# ADVECTION DOMINATED ACCRETION FLOWS AROUND KERR BLACK HOLES

M. A. Abramowicz[1,2], X.-M. Chen[1,3], M. Granath [1], and

J.-P. Lasota[4,5,6]

## ABSTRACT

We derive all relevant equations needed for constructing a global general relativistic model of advectively cooled, very hot, optically thin accretion disks around black holes and present solutions which describe advection dominated flows in the gravitational field of a Kerr black hole.

*Subject headings:* accretion, accretion disks — black hole physics — hydrodynamics — radiative mechanisms — relativity

## 1. ADVECTION DOMINATED BLACK HOLE ACCRETION DISKS

Recently, solutions corresponding to a new type of hot, optically thin accretion disks cooled mostly by advection has been discovered (see Abramowicz & Lasota 1996, Abramowicz 1996, and Narayan 1996 for reviews). The luminosity, $L$, of these disks is much lower than that of the standard disks with the same rate of mass supply $\dot M$. The radiative cooling efficiency, defined as $\eta = L/\dot M c^2$, is typically less than 0.01 in the newly discovered disks, while in the case of the standard disks it is typically $\eta \gtrsim 0.1$. Thus, the advection dominated accretion flows (ADAFs) hide the evidence that they are consuming a lot of matter — for this reason Abramowicz & Lasota (1996) called them "secret guzzlers". A typical secret guzzler appears to be very hot and underluminous.

[1]Department of Astronomy and Astrophysics, Göteborg University and Chalmers University of Technology, 412 96 Göteborg, Sweden

[2]Nordita, Blegdamsvej 17, DK-2100, Copenhagen Ø, Denmark

[3]Department of Physics and Astronomy, Northwestern University, Evanston, IL 60208

[4]Belkin Visiting Professor, Department of Condensed Matter Physics, Weizmann Institute of Science, 76100 Rehovot, Israel

[5]Lady Davis Visiting Professor, Department of Physics, Technion, 3200 Haifa, Israel

[6]UPR 176 du CNRS; DARC, Observatoire de Paris, Section de Meudon, 92195 Meudon, France

Lasota, Narayan and collaborators have discussed properties of the secret guzzler models and compared them with observed properties of sources which are believed to contain accreting black holes and appear both underluminous and very hot, like Sgr A* (Narayan et al. 1995), soft X-ray transients (Narayan et al. 1996b, Lasota et al. 1996a), NGC 4258 (Lasota et al. 1996b). These detailed discussions, based on calculations of electromagnetic spectra, convincingly demonstrated that secret guzzlers may explain these particular types of sources, with no need for ad hoc hypothesis about additional optically thin components in the model, or some unusual properties of the viscosity.

Models of secret guzzlers are described by similar equations as the well-known slim accretion disks (Abramowicz et al. 1988). The only difference is in equations for the radiative processes: secret guzzlers are hot and optically thin and the ion temperature is, most probably, much higher than that of the electrons (see e.g. Narayan 1996) and slim disks are cool and optically thick. All the previously calculated models of slim-disks and secret guzzlers are not fully satisfactory at least in one respect: they assume an approximate model for the gravity of the black hole. It is either just the plain Newton's potential of a point mass $M$, $\Phi = -GM/r$, or the Paczyński's pseudo-potential $\Phi = -GM/(r - r_G)$, $r_G = 2GM/c^2$ (Paczyński & Wiita, 1980, hereafter PW). No general relativistic solution has ever been obtained.

The PW paradigm is an *excellent* approximation for the calculations of the hydrodynamical properties of accretion disk around non-rotating black holes. It is both very simple and very accurate. However, it has two shortcomings: (1) it does not describe the relativistic effects in the light propagation, and (2) it does not include the effect of the inertial frame dragging, an important property of rotating black holes. For this reason, at the TAD2 meeting at Garching (see Duschl at al. 1994) one of us suggested that the time has come to construct solutions for slim disks that would be *fully consistent* with general relativity. The first step should be a derivation of the relevant equations. Ramesh Narayan argued that it should be the job of the proposer so this was done in Lasota (1994; thereafter L94).

We repeat this calculations here, correcting some computational errors of L94 and using a slightly more practical notation. The general assumption and definition are presented in Section 2. In Section 3 we give the set of equations describing a general relativistic model of a hot accretion flow. The detailed microphysics that describes the radiative cooling is briefed in Section 4 and 5. After a discussion of the transonic character of the flow in Section 6, numerical solutions of advection dominated flows in the gravitational field of a Kerr black hole are presented in Section 7. Section 8 contains the conclusions.

## 2. ASSUMPTIONS AND DEFINITIONS

We use geometrical units here that are linked to the physical units for length, time and mass by,

$$\begin{pmatrix} \text{length} \\ \text{in physical units} \end{pmatrix} = \begin{pmatrix} \text{length} \\ \text{in geometrical units} \end{pmatrix}, \quad (1)$$

$$\begin{pmatrix} \text{time} \\ \text{in physical units} \end{pmatrix} = \frac{1}{c} \begin{pmatrix} \text{time} \\ \text{in geometrical units} \end{pmatrix}, \quad (2)$$

$$\begin{pmatrix} \text{mass} \\ \text{in physical units} \end{pmatrix} = \frac{c^2}{G} \begin{pmatrix} \text{mass} \\ \text{in geometrical units} \end{pmatrix}. \quad (3)$$

We use the Boyer-Lindquist spherical coordinates $t, r, \theta, \varphi$ to describe the Kerr black hole metric. The cylindrical vertical coordinate $z = \cos\theta$ is defined very close to the equatorial plane, $z = 0$. The metric of the Kerr black hole on the equatorial plane, accurate up to the $(z/r)^0$ terms, takes the form given by Novikov & Thorne (1973, hereafter NT),

$$ds^2 = -\frac{r^2 \Delta}{A} dt^2 + \frac{A}{r^2} (d\varphi - \omega dt)^2 + \frac{r^2}{\Delta} dr^2 + dz^2 \quad (4)$$

$$\Delta = r^2 - 2Mr + a^2, \quad A = r^4 + r^2 a^2 + 2Mra^2, \quad \omega = \frac{2Mar}{A}. \quad (5)$$

Here $M$ is the mass, and $a$ the total specific angular of the Kerr black hole. The metric (4) was also used in L94. Note, that the determinant of the metric tensor corresponding to (4) is $g = -r^2$.

There are two Killing vectors in the geometry (4), connected to the time symmetry (metric is independent of $t$), axial symmetry (metric is independent on $\varphi$)

$$\eta^i = \delta^i_{(t)}, \quad \xi^i = \delta^i_{(\varphi)}, \quad (6)$$

where $\delta^i_{(k)}$ is the Kronecker delta. Using Killing vectors (6) one defines some useful scalar functions: angular velocity of the dragging of inertial frames $\omega$, gravitational potential $\Phi$, and radius of gyration $\tilde{R}$,

$$\omega = -\frac{(\eta\xi)}{(\xi\xi)}, \quad e^{-2\Phi} = (\eta\eta) - \omega^2(\xi\xi), \quad \tilde{R}^2 = (\xi\xi)e^{2\Phi}. \quad (7)$$

In the Boyer-Lindquist coordinates the scalar products of the Killing vectors (6) are given by the components of the metric,

$$(\eta\eta) = g_{tt}, \quad (\eta\xi) = g_{t\varphi}, \quad (\xi\xi) = g_{\varphi\varphi}, \quad (8)$$

and therefore quantities defined by (7) can by explicitly written down in terms of the Boyer-Lindquist coordinates:

$$\tilde{R}^2 = \frac{A^2}{r^4 \Delta}, \quad e^{-2\Phi} = \frac{r^2 \Delta}{A}. \tag{9}$$

The black hole surface (event horizon) is at

$$r_+ = M + \left(M^2 - a^2\right)^{1/2}. \tag{10}$$

The unit timelike vector

$$n^i = e^{\Phi} \left(\eta^i + \omega \xi^i\right), \tag{11}$$

is orthogonal to the space-like surfaces $t$ =const. It corresponds to the four-velocity of the local inertial observer or ZAMO, i.e. Zero Angular Momentum Observers (Bardeen 1973).

The four velocity of matter $u^i$ has components $u^t$, $u^\varphi$, $u^r$,

$$u^i = u^t \delta^i_{(t)} + u^\varphi \delta^i_{(\varphi)} + u^r \delta^i_{(r)}. \tag{12}$$

One defines the angular velocity $\Omega$ with respect to the stationary observer, and the angular velocity $\tilde{\Omega}$ with respect to the local inertial observer by,

$$\Omega = \frac{u^\varphi}{u^t}, \quad \tilde{\Omega} = \Omega - \omega, \tag{13}$$

The angular frequencies of the corotating (+) and counterrotating (−) Keplerian orbits are

$$\Omega_K^\pm = \pm \frac{M^{1/2}}{r^{3/2} \pm a M^{1/2}}, \tag{14}$$

and the Keplerian specific angular momentum is given by

$$\mathcal{L}_K^\pm = \pm \frac{M^{1/2} \left(r^2 \mp 2aM^{1/2} r^{1/2} + a^2\right)}{r^{3/4} \left(r^{3/2} - 3Mr^{1/2} \pm 2aM^{1/2}\right)^{1/2}}. \tag{15}$$

The Keplerian angular momentum has a minimum at the marginally stable orbit

$$\begin{aligned} r_{ms}^\pm &= M\{3 + Z_2 \mp [(3 - Z_1)(3 + Z_1 + 2Z_2)]^{1/2}\}, \\ Z_1 &= 1 + \left(1 - a^2/M^2\right)^{1/3} \left[(1 + a/M)^{1/3} + (1 - a/M)^{1/3}\right], \\ Z_2 &= \left(3a^2/M^2 + Z_1^2\right)^{1/2}. \end{aligned} \tag{16}$$

In the reference frame of the local inertial observer the four velocity takes the form,

$$u^i = \gamma \left(n^i + v^{(\varphi)} \tau^i_{(\varphi)} + v^{(r)} \tau^i_{(r)}\right). \tag{17}$$

The vectors $\tau^i_{(\varphi)}$ and $\tau^i_{(r)}$ are the unit vectors in the coordinate directions $\varphi$ and $r$. The Lorentz gamma factor $\gamma$ equals,

$$\gamma = \frac{1}{\sqrt{1 - \left(v^{(\varphi)}\right)^2 - \left(v^{(r)}\right)^2}}. \tag{18}$$

The relation between the Boyer-Lindquist and the physical velocity component in the azimuthal direction is,

$$v^{(\varphi)} = \tilde{R}\tilde{\Omega}. \tag{19}$$

It will be convenient to use the (rescaled) radial velocity component $V$ defined by the formula,

$$\frac{V}{\sqrt{1-V^2}} = \gamma v^{(r)} = u^r g_{rr}^{1/2}. \tag{20}$$

The Lorentz gamma factor may be written as,

$$\gamma^2 = \left(\frac{1}{1 - \tilde{\Omega}^2 \tilde{R}^2}\right)\left(\frac{1}{1-V^2}\right), \tag{21}$$

and from this follows a simple expression for $V$ in terms of the velocity components measured in the frame of the local inertial observer,

$$V = \frac{v^{(r)}}{\sqrt{1 - \left(v^{(\varphi)}\right)^2}} = \frac{v^{(r)}}{\sqrt{1 - \tilde{R}^2\tilde{\Omega}^2}}. \tag{22}$$

Thus, $V$ is the radial velocity of the fluid as measured by an observer at fixed $r$ who corotates with the fluid. In the notation used in L94 (cf. his equation [17]),

$$v^{(r)} = \left[v^{\hat{r}}\right]_{\text{L94}}. \tag{23}$$

Although a different quantity could have been chosen as the definition of the "radial velocity", only $V$ has the three very convenient properties, all guaranteed by its definition: (i) everywhere in the flow $|V| \leq 1$, (ii) on the horizon $|V| = 1$, (iii) at the sonic point $|V| \approx c_s$, with $c_s$ being the local sound speed. To see that the first property holds, let us define $\tilde{V}^2 = u^r u_r = u^r u^r g_{rr} \geq 0$. Then, one has $V^2 = \tilde{V}^2/(1 + \tilde{V}^2) \leq 1$. Writing $V = 1/\sqrt{1 + \Delta/(r^2 u^r u^r)}$ demonstrates property (ii). Property (iii) of $V$ will be proved in Section 6. Other possible choices of the "radial velocity" are not that convenient. For example, the radial velocity $u = |u^r|$, chosen by Shapiro & Teukolsky, (1983) has none of these properties and would be less convenient in the present context.

The stress-energy tensor $T^{ik}$ of the matter in the disk is given by,

$$T^{ik} = (\varepsilon + p)\, u^i u^k + p\, g^{ik} + S^{ik} + u^k q^i + u^i q^k, \tag{24}$$

where $\varepsilon$ is the total energy density, $p$ is the pressure,

$$S_{ik} = \nu \rho \sigma_{ik}, \tag{25}$$

is the viscous stress tensor, $\rho$ is the ress mass density and $q^i$ is the radiative energy flux. In the last equation $\nu$ is the kinematic viscosity coefficient and $\sigma_{ik}$ is the shear tensor of the velocity field. From the first law of thermodynamic it follows that

$$d\varepsilon = \frac{\varepsilon + p}{\rho} d\rho + \rho T dS, \tag{26}$$

where $T$ is the temperature and $S$ is the entropy per unit mass. Note, that in the physical units that $\varepsilon = \rho c^2 + \Pi$, where $\Pi$ is the internal energy. For non-relativistic fluids, $\Pi \ll \rho c^2$ and $p \ll \rho c^2$, and therefore

$$\varepsilon + p \approx c^2 \rho. \tag{27}$$

We shall use this approximation (in geometrical units $\varepsilon + p \approx \rho$) in all our calculations. This approximation does not automatically ensure that the sound speed is below $c$, and one should check this a posteriori when models are constructed. We write the first law of thermodynamics in the form:

$$dU = -p \left(\frac{1}{\rho}\right) + TdS \tag{28}$$

where $U = \Pi/\rho$.

## 3. SLIM DISK EQUATIONS IN KERR GEOMETRY

It is convenient to write the final form of all the slim disk equations at the equatorial plane, $z = 0$. Only these equations which do not refer to the vertical structure could be derived directly from the quantities at the equatorial plane with no further approximations. All other equations are approximated — either by expansion in terms of the relative disk thickness $H/r$, or by vertical averaging. The thickness of the disk $H$ is defined as in equation (44).

### 3.1. Mass conservation equation

From general equation of mass conservation,

$$\nabla^i (\rho u_i) = 0, \tag{29}$$

and definition of the surface density $\Sigma$,

$$\Sigma = \int_{-H(r)}^{+H(r)} \rho(r, z) dz \approx 2H\rho, \tag{30}$$

we derive the mass conservation equation,

$$\dot{M} = -2\pi \Delta^{1/2} \Sigma \frac{V}{\sqrt{1 - V^2}}. \tag{31}$$

It is identical to that derived by NT. In L94 the numerical factor on the right hand side is $-4\pi$.

### 3.2. Equation of angular momentum conservation

From the general form of the angular momentum conservation,

$$\nabla_k \left( T^{ki} \xi_i \right) = 0, \tag{32}$$

we derive, after some algebra,

$$\frac{\dot{M}}{2\pi r} \frac{d\mathcal{L}}{dr} + \frac{1}{r} \frac{d}{dr} \left( \Sigma \nu A^{3/2} \frac{\Delta^{1/2} \gamma^3}{r^4} \frac{d\Omega}{dr} \right) - F^- \mathcal{L} = 0, \tag{33}$$

where $F^- = 2q_z$ is the vertical flux of radiation, and

$$\mathcal{L} \equiv -(u\xi) = -u_\varphi = \gamma \left( \frac{A^{3/2}}{r^3 \Delta^{1/2}} \right) \tilde{\Omega}, \tag{34}$$

is the specific (per unit mass) angular momentum. Equation (33) differs from that derived in L94. In addition to a trivial misprint in the L94 equation, there is also a more significant difference — we keep the term $F^- \mathcal{L}$ that was rejected from the final form of the L94 equation. This term represents angular momentum losses through radiation. Although it was always fully recognized that angular momentum may be lost this way, it has been argued that this term must be very small. Rejection of this term would enormously simplify numerical calculations, because it is obvious that with $F^- \mathcal{L} = 0$ equation (33) can be trivially integrated,

$$\frac{\dot{M}}{2\pi} (\mathcal{L} - \mathcal{L}_0) = -\Sigma \nu A^{3/2} \frac{\Delta^{1/2} \gamma^3}{r^4} \frac{d\Omega}{dr}, \tag{35}$$

where $\mathcal{L}_0$ is the specific angular momentum of matter at the horizon ($\Delta = 0$). In the numerical scheme for integrating the slim Kerr equations (with $F^- \mathcal{L}$ assumed to be zero) the quantity $\mathcal{L}_0$ plays an important role: it is the eigenvalue of the solutions that passes regularly through the sonic point.

Several authors have pointed out that in some astrophysically relevant situations the $F^- \mathcal{L}$ cannot be put to zero, because it could be a significant part (up to 0.4) of the angular momentum balance (see Lamb, 1996 for discussion and most recent references).

### 3.3. Equation of momentum conservation

From the r-component of the equation $\nabla_i T^{ik} = 0$ one derives

$$\frac{V}{1-V^2} \frac{dV}{dr} = \frac{\mathcal{A}}{r} - \frac{1}{\Sigma} \frac{dP}{dr}, \tag{36}$$

where $P = 2Hp$ is the vertically integrated pressure and

$$\mathcal{A} = -\frac{MA}{r^3 \Delta \Omega_k^+ \Omega_k^-} \frac{(\Omega - \Omega_k^+)(\Omega - \Omega_k^-)}{1 - \tilde{\Omega}^2 \tilde{R}^2}. \tag{37}$$

L94 version of the momentum conservations equation contains few trivial misprints.

Using equation (20) one shows that (36) has the correct limit corresponding to the spherical accretion ($\Omega = 0$) on the non-rotating ($a = 0$) black hole: in this limit it becomes identical with equation (G.7) on page 569 of Shapiro & Teukolsky (1983).

### 3.4. Equation of energy conservation

From the general form of the energy conservation

$$\nabla_i \left(T^{ik} \eta_k\right) = 0, \tag{38}$$

and the first law of thermodynamics,

$$T = \frac{1}{\rho}\left(\frac{\partial \varepsilon}{\partial S}\right)_\rho, \quad p = \rho\left(\frac{\partial \varepsilon}{\partial \rho}\right)_S - \varepsilon, \tag{39}$$

the energy equation can be written in general as

$$F^{\mathrm{adv}} = F^+ - F^-, \tag{40}$$

where

$$F^+ = \nu \Sigma \frac{A^2}{r^6} \gamma^4 \left(\frac{d\Omega}{dr}\right)^2 \tag{41}$$

is the surface heat generation rate, $F^-$ is the radiative cooling flux (both surface) which is discussed in Section 5, and $F^{\mathrm{adv}}$ is the advective cooling rate due to the radial motion of the gas. It is expressed as

$$F^{\mathrm{adv}} = \frac{\Sigma V}{\sqrt{1 - V^2}} \frac{\Delta^{1/2}}{r} T \frac{dS}{dr} \equiv -\frac{\dot{M}}{2\pi r} T \frac{dS}{dr}. \tag{42}$$

### 3.5. Equation of vertical balance of forces

One should calculate the condition for the hydrostatic vertical equilibrium in the frame that comoves with matter. NT have approximated this frame by the one which moves along circular Keplerian orbits at the equatorial plane. It is known that in the innermost part of the disk the motion of matter differs considerably from the circular Keplerian motion. For this reason, one needs to repeat the NT calculations in the comoving frame of matter. [7]

Like NT, we demand that the vertical pressure force should be equal to the vertical component of the tidal gravitational force,

$$\frac{1}{\rho}\left(\frac{\partial p}{\partial z}\right)_H = H R^i{}_{jkl} \tau^m_{(z)} g_{mi} \tau^k_{(z)} u^j u^l \approx H\gamma^2 R^i{}_{jkl} \tau^m_{(z)} g_{mi} \tau^k_{(z)} n^j n^l = H\gamma^2 R^{(z)}_{(t)(z)(t)}, \qquad (43)$$

however we demand this to happen in the comoving frame of the fluid (note the fluid's four velocity $u^j u^l$ multyplying the Riemann tensor $R^i{}_{jkl}$), while NT demand that equality of forces should occur in the frame rotating with the Keplerian velocity (NT have $u^j_K u^l_K$ multiplying the Riemann tensor). In the formula $\tau^k_{(z)}$ is the unit vector in the $z$-direction. From this we derive the final formula,

$$\frac{p}{\rho H^2} = \gamma^2 \frac{M}{r^3} \left[\frac{(r^2+a^2)^2 + 2\Delta a^2}{(r^2+a^2)^2 - \Delta a^2}\right] = \gamma^2 \mathcal{G}, \qquad (44)$$

that should be compared with NT formula (5.7.2). The difference is that in our formula the redshift factor $\gamma$ is given by (21), which corresponds to the motion of matter and not to a fictitious fluid with strictly Keplerian rotation. The NT formula is singular at the location of the photon circular trajectory, $r = r_{ph}$, and our formula at the horizon. L94 gives the same formula for the vertical equilibrium.

The singularity of Eq. (44) on the horizon is an artifact of approximations used in Eq. (43). We assumed that the flow is vertical hydrostatic equilibrium, in the ZAMO frame, down to the black hole surface. This requirement implies an infinite boost at the horizon. In reality the flow near the horizon is supersonic and therefore practically in a free fall. Eq. (43) should be completed by a term $v^r (\partial v^z/\partial r)$ which, in the supersonic flow, will dominate over the pressure gradient term. As a result Eq. (43) will become the (non-singular) geodesic deviation equation. However, because we are adopting here a 1D approach in which 'vertical' quantities are averaged it is not possible to include $v_z$ terms in the equations. Since the 'singularity' appears only in the supersonic part of the transonic flow and since the reason for its appearance is clear, it has no influence on the physically important properties of the accretion flow (see also discussion in Section 6).

### 3.6. Viscosity prescription

---

[7]This reason differs from that discussed recently by Riffert & Herold (1995, Ap.J., 450, 508) who have also re-derived some of the NT formulae. They have argued that for some of these derivations, the Kerr metric must be accurate to second order in $z/r$. Although they have kept the $(z/r)^2$ terms in geometry, they have rejected physical quantities of this order: radial pressure gradient, radial advective cooling, etc., which is not a self-consistent scheme.

The standard assumption for the viscosity coefficient is,

$$\nu = (2/3)\alpha c_s H, \tag{45}$$

where $c_s = \sqrt{p/\rho}$ is the isothermal sound speed.

## 4. THERMODYNAMICAL RELATIONS

The equation of state can be expressed in the form:

$$p = p_r + \frac{\mathcal{R}}{\mu_i}\rho T_i + \frac{\mathcal{R}}{\mu_e}\rho T_e + \frac{B^2}{24\pi}, \tag{46}$$

where $p_r$ is the radiation pressure, $\mathcal{R}$ is the gas constant, $\mu_i$ and $\mu_e$ the mean molecular weights of ions and electrons respectively, $T_i$, and $T_e$ ion and electron temperatures, $a$ the radiation constant, and $B$ the intensity of a isotropically tangled magnetic field, includes the radiation, gas and magnetic pressures. The radiation pressure $p_r$, the gas pressure $p_g$, and the magnetic pressure $p_m$ correspond respectively to the first term, the second and third terms, and the last term in equation (46).

The mean molecular weights of ions and electrons can be well approximated by:

$$\mu_i \approx \frac{4}{4X+Y}, \qquad \mu_i \approx \frac{2}{1+X}, \tag{47}$$

where $X$ is the relative mass abundance of hydrogen and $Y$ that of helium. We may define a temperature as

$$T = \mu\left(\frac{T_i}{\mu_i} + \frac{T_e}{\mu_e}\right), \tag{48}$$

where

$$\mu = \left(\frac{1}{\mu_i} + \frac{1}{\mu_e}\right)^{-1} \approx \frac{2}{1+3X+1/2Y} \tag{49}$$

is the mean molecular weight in the standard approximation (Cox & Giuli 1968). In the case of a one-temperature gas ($T_i = T_e$), one has $T = T_i = T_e$. For an optically thick gas, $p_r = \frac{1}{3}aT_r^4$.

Until the end of this section we will assume that $p_r = \frac{1}{3}aT_r^4$. As noted by Narayan & Yi (1995b; hereafter NY), the frozen-in magnetic field pressure $p_m \sim B^2 \sim \rho^{4/3}$, therefore we may write the internal energy as

$$U = \frac{aT_r^4}{\rho} + \frac{\mathcal{R}T}{\mu m_u(\gamma_g - 1)} + e_o\rho^{1/3}, \tag{50}$$

where $e_o$ is a constant ($p_m = 1/3e_o\rho^{4/3}$) and $\gamma_g$ is the ratio of the specific heats of the gas. We define

$$\beta = \frac{p_g}{p}, \quad \beta_m = \frac{p_g}{p_g + p_m}, \quad \beta^* = \frac{4 - \beta_m}{3\beta_m}\beta. \tag{51}$$

From equations (46) and (50) one obtains the following formulae (see e.g. Cox & Giuli 1968) for the specific heat at constant volume:

$$c_V = \frac{\mathcal{R}}{\mu(\gamma_g - 1)}\left[\frac{12(1 - \beta/\beta_m)(\gamma_g - 1) + \beta}{\beta}\right] = \frac{4 - 3\beta^*}{\Gamma_3 - 1}\frac{p}{\rho T} \tag{52}$$

and the adiabatic indices:

$$\Gamma_3 - 1 = \frac{(4 - 3\beta^*)(\gamma_g - 1)}{12(1 - \beta/\beta_m)(\gamma_g - 1) + \beta} \tag{53}$$

$$\Gamma_1 = \beta^* + (4 - 3\beta^*)(\Gamma_3 - 1). \tag{54}$$

The ratio of specific heats is $\gamma = c_p/c_V = \Gamma_1/\beta$. For $\beta = \beta_m$ we have $\Gamma_3 = \gamma_g$ and $\Gamma_1 = (4 - \beta)/3 + \beta(\gamma_g - 1)$. For an equipartition magnetic field ($\beta = 0.5$) one gets $\Gamma_1 = 1.5$ and for $\beta = 0.95$, $\Gamma_1 = 1.65$ (here we have used $\gamma_g = 5/3$). Our formula for $\Gamma_1$ is different from that used by NY but the numerical values differ by less than 5%. One expects $\beta_m \sim 0.5 - 1$. Since

$$T\frac{dS}{dr} = c_V\left[\frac{d\ln T}{dr} - (\Gamma_3 - 1)\left(\frac{d\ln \Sigma}{dr} - \frac{d\ln H}{dr}\right)\right], \tag{55}$$

In the numerical scheme based on Chen & Taam (1993) (see also Abramowicz et al 1995) the advective flux is written in the form:

$$F^{\text{adv}} = \frac{\dot{M}}{2\pi r^2}\frac{p}{\rho}\xi_a \tag{56}$$

where

$$\xi_a = -\left[\frac{4 - 3\beta^*}{\Gamma_3 - 1}\frac{d\ln T}{d\ln r} + (4 - 3\beta^*)\frac{d\ln \Sigma}{d\ln r}\right]. \tag{57}$$

Here the term $\propto d\ln H/d\ln r$ is neglected. Since no rigorous vertical averaging procedure exists, the presence or not of the $d\ln H/d\ln r$ – type terms in this (and other) equation may be decided only by comparison with 2D calculations (see e.g. Narayan & Yi 1995a and also Chen et al 1995 for a discussion of this point).

For $\beta_m = 1$ we recover the formulae used in Abramowicz et al (1995). The formulae derived in this section are valid for the optically thin case $\tau = 0$ if one assumes $\beta = \beta_m$.

## 5. RADIATIVE COOLING

We have to complete the set of equations by specifying the physical processes which will be involved in the radiative cooling. In our opinion the most convenient and general description of cooling processes have been presented by NY. They consider a two-temperature plasma cooled by synchrotron radiation, inverse Compton process and bremsstrahlung emission. They neglect electron–positron pair creation and annihilation but as shown by Björnsson et al (1996) and Kusunose & Mineshige (1996) this is justified in most cases of interest. Below, for the reader's convenience we will recall the formulae used by NY.

### 5.1. Heating of electrons by ions

NY use the formula of Stepney & Guilbert (1983) which has been modified in order to account for a $X = 0.75$ and $Y = 0.25$ composition (the effective molecular weight of ions $\mu_i = 1.23$, of electrons $\mu_e = 1.14$). Here we give the general formula in which the factor $n_i$ in the Stepney & Guilbert (1983) formula is replaced by,

$$\bar{n} = \sum Z_j^2 n_j, \tag{58}$$

where $Z_j$ and $n_j$ are respectively the charge and the number density of $j$th species. The Coulomb collisions transfer energy from (hotter) ions to electrons at a volume transfer rate

$$\begin{aligned} f^{ie} &= \frac{3}{2}\frac{m_e}{m_i} n_e \bar{n} \sigma_T c \frac{kT_i - kT_e}{K_2(1/\theta_e) K_2(1/\theta_i)} \ln \Lambda \\ &\times \left[ \frac{2(\theta_e + \theta_i)^2 + 1}{\theta_e + \theta_i} K_1\left(\frac{\theta_e + \theta_i}{\theta_e \theta_i}\right) + 2 K_0\left(\frac{\theta_e + \theta_i}{\theta_e \theta_i}\right) \right] \text{ erg cm}^{-3}\text{ s}^{-1} \end{aligned} \tag{59}$$

where the $K$'s are modified Bessel functions $\ln \Lambda \approx 20$ is the Coulomb logarithm and the dimensionless electron and ion temperatures are defined by

$$\theta_e = kT_e/m_e c^2, \qquad \theta_i = kT_i/m_e c^2. \tag{60}$$

### 5.2. Bremsstrahlung Cooling

The bremsstrahlung includes emission from both ion-electron and electron-electron collisions (Stepney & Guilbert 1983; Svensson 1982; NY)

$$f_{\text{br}}^- = f_{ei}^- + f_{ee}^-. \tag{61}$$

The ion-electron bremsstrahlung cooling is given by

$$f_{ei}^- = n_e \bar{n} \sigma_T c \alpha_f m_e c^2 F_{ei}(\theta_e) \tag{62}$$

where $\alpha_f = 1/137$ is the fine structure constant and

$$
\begin{aligned}
F_{ei}(\theta_e) &= 4\left(\frac{20\theta_e}{\pi^3}\right)^{1/2}(1+1.781\theta_e^{1.34}), \quad \theta_e < 1, \\
&= \frac{90}{2\pi}[\ln(1.123\theta_e + 0.48) + 1.5], \quad \theta_e > 1.
\end{aligned}
\quad (63)
$$

In the original formula quoted by Stepney & Guilbert (1983) there is a number 0.42 instead of 0.48 (see NY).

In the non-relativistic approximation one gets from Eq. (62)

$$
f_{ei}^- = 1.57 \times 10^{-27} n_e \bar{n} T_e^{1/2} \quad \text{erg cm}^{-3} \text{ s}^{-1}, \quad (64)
$$

which for Population I abundances ($X = 0.7$, $Y = 0.28$) gives

$$
f_{ei}^- = 5.63 \times 10^{20} \rho^2 T_e^{1/2} \quad \text{erg cm}^{-3} \text{ s}^{-1}. \quad (65)
$$

For the electron–electron bremsstrahlung, Svensson (1982) gives the following formula for $\theta_e < 1$:

$$
\begin{aligned}
f_{ee}^- &= n_e^2 c r_e^2 m_e c^2 \alpha_f \frac{20}{9\pi^{1/2}}(44 - 3\pi^2)\theta_e^{3/2} \\
&\times (1 + 1.1\theta_e^2 - 1.25\theta_e^{5/2}) \quad \text{erg cm}^{-3} \text{ s}^{-1},
\end{aligned}
\quad (66)
$$

and for $\theta_e > 1$:

$$
f_{ee}^- = n_e^2 c r_e^2 m_e c^2 \alpha_f 24 \theta_e [\ln(\eta\theta_e) + 1.28] \quad \text{erg cm}^{-3} \text{ s}^{-1}, \quad (67)
$$

where $r_e = e^2/m_e c^2$ is the classical radius of electron and $\eta = \exp(-\gamma_E) = 0.5616$. Again, in the original formula there is a 5/4 instead of 1.28 (see NY).

### 5.3. Synchrotron cooling

Following NY we give formulae for the synchrotron emission of a relativistic Maxwellian distribution of electrons. These formulae are valid only for $\theta > 1$ but it is sufficient for applications to ADAFs (see NY).

$$
f_{\text{synch}}^- = \frac{2\pi}{3c^2} k T_e(r) \frac{d\nu_c^3(r)}{dr}, \quad (68)
$$

where

$$
\nu_c = \frac{3}{2}\frac{eB}{2\pi m_e c}\theta_e^2 x_M, \quad (69)
$$

and $x_M$ is the solution of the transcendental equation

$$\exp(1.8899 x_M^{1/3}) = 2.49 \times 10^{-10} \frac{4\pi n_e r}{B} \frac{1}{\theta_e^3 K_2(1/\theta_e)}$$
$$\times \left( \frac{1}{x_M^{7/6}} + \frac{0.40}{x_M^{17/12}} + \frac{0.5316}{x_M^{5/3}} \right), \tag{70}$$

where radius $r$ is in physical unit of cm.

### 5.4. Compton cooling

NY use the Dermer, Liang & Canfield (1991) prescription for the Compton energy enhancement factor $\eta$:

$$\eta = 1 + \frac{p(A-1)}{1-PA} \left[ 1 - \left( \frac{x}{3\theta_e} \right)^{-1-\ln P/\ln A} \right]$$
$$\equiv 1 + \eta_1 + \eta_2 \left( \frac{x}{\theta_e} \right)^{\eta_3}, \tag{71}$$

where

$$x = \frac{h\nu}{m_e c^2},$$
$$P = 1 - \exp(-\tau_{es}),$$
$$A = 1 + 4\theta_e + 16\theta_e^2,$$
$$\eta_1 = \frac{p(A-1)}{1-PA},$$
$$\eta_2 = -3^{-\eta_3} \eta_1,$$
$$\eta_3 = -1 - \ln P/\ln A, \tag{72}$$

and $\tau_{es}$ is the electron scattering optical depth. Note that $\eta$, $P$, $A$ here are different from the ones defined in previous sections.

The Comptonization of the bremsstrahlung radiation is given by

$$f_{\mathrm{br,C}}^- = 3\eta_1 q_{br}^- \left\{ \left( \frac{1}{3} - \frac{x_c}{3\theta_e} \right) - \frac{1}{\eta_3 + 1} \left[ \left( \frac{1}{3} \right)^{\eta_3+1} - \left( \frac{1}{3\theta_e} \right)^{\eta_3+1} \right] \right\}, \tag{73}$$

and the Comptonized synchrotron emission is

$$f_{\mathrm{synch,C}}^- = f_{\mathrm{synch}}^- \left[ \eta_1 - \eta_2 (x_c/\theta_e)^{\eta_3} \right], \tag{74}$$

where $x_c = \nu_c / m_e c^2$.

## 5.5. Total radiative cooling

In the optically thin case the total radiative cooling will be given by

$$f^- = \frac{F^-}{2H} = f^-_{\text{br}} + f^-_{\text{synch}} + f^-_{\text{br,C}} + f^-_{\text{synch,C}}, \tag{75}$$

whereas in the optically thick case one can use

$$f^- = \frac{8\sigma T_e^4}{3H\tau} \tag{76}$$

where $\tau = \tau_{\text{abs}} + \tau_{\text{es}} = 0.5(\kappa_{abs} + \kappa_{es})\Sigma$ is the total optical depth. In the intermediate case one should solve the transfer equation to get reliable results. In the present context one can use the solution of the grey problem obtained by Hubeny (1990). NY give the following formula:

$$f^- = \frac{4\sigma T_e^4}{H} \left[ \frac{3\tau}{2} + \sqrt{3} + \frac{4\sigma T_e^4}{H} \times (f^-_{\text{br}} + f^-_{\text{synch}} + f^-_{\text{br,C}} + f^-_{\text{synch,C}})^{-1} \right]^{-1}, \tag{77}$$

where the relation:

$$\tau_{\text{abs}} = \frac{H}{4\sigma T_e^4}(f^-_{\text{br}} + f^-_{\text{synch}} + f^-_{\text{br,C}} + f^-_{\text{synch,C}}) \tag{78}$$

has been used. The opacity appearing in Eq. (78) is the *Planck mean* opacity.

The radiation pressure can be expressed as

$$p_r = \frac{f^- H}{2c}\left(\tau + \frac{2}{\sqrt{3}}\right). \tag{79}$$

## 5.6. Thermal balance equations

Since the viscous heating concerns the ions we shall write equation (40) in the form of two equations (NY):

$$\begin{aligned} f^+ &= f^{\text{adv}} + f^{ie}, \\ f^{ie} &= f^-, \end{aligned} \tag{80}$$

where $f^+ = F^+/(2H)$, $f^{\text{adv}} = F^{\text{adv}}/(2H)$, $f^{ie}$ and $f^-$ are given by equations (59) and (77) respectively.

## 6. THE SONIC POINT AND THE BOUNDARY CONDITIONS

Our system of equations, like its Newtonian and PW versions, possesses singular points. In this section we will use the vertically averaged quantities $P$ and $\Sigma$. From equation (31) we have

$$\frac{d\ln\Sigma}{d\ln r} = -\frac{1}{V(1-V^2)}\frac{dV}{d\ln r} - \mathcal{B}, \tag{81}$$

where,

$$\mathcal{B} = \frac{d\ln(\Delta^{1/2})}{d\ln r} = \frac{r(r-M)}{\Delta}. \tag{82}$$

The radial gradient of $P$ in equation (36) can be derived from equation (44) and (46) as

$$\frac{d\ln P}{d\ln r} = \frac{2(4-3\beta^*)}{1+\beta^*}\frac{d\ln T}{d\ln r} + \frac{3\beta^*-1}{1+\beta^*}\frac{d\ln\Sigma}{d\ln r} - \frac{(1-\beta^*)}{1+\beta^*}\frac{d\ln\gamma^2\mathcal{G}}{d\ln r}. \tag{83}$$

Thus, by combining the radial motion equation (36) and the energy equation (40), one gets

$$\frac{d\ln V}{d\ln r} = \frac{\mathcal{N}}{\mathcal{D}}(1-V^2), \tag{84}$$

where

$$\mathcal{D} = \mathcal{C}c_s^2 - \left[1 + \frac{2(1-\beta^*)}{1+\beta^*}\right]V^2, \tag{85}$$

$$\mathcal{N} = -\mathcal{A} - \mathcal{B}c_s^2\left[\mathcal{C} - \frac{2(1-\beta^*)}{1+\beta^*}\right] - \frac{(\Gamma_3-1)4\pi r^2}{(1+\beta^*)\dot{M}}(F^+ - F^-) \\ - \frac{(1-\beta^*)}{(1+\beta^*)}c_s^2\left[\frac{d\ln\mathcal{G}}{d\ln r} + \frac{2v^{(\varphi)}\left(dv^{(\varphi)}/dr\right)}{1-\left(v^{(\varphi)}\right)^2} + 2\mathcal{B}\right]. \tag{86}$$

$\mathcal{A}$ is defined by Eq. (37) and

$$c_s^2 = \frac{P}{\Sigma}, \quad \mathcal{C} = 1 + 2\frac{\Gamma_1 - 1}{1+\beta^*}. \tag{87}$$

The equation for the temperature is then

$$\frac{d\ln T}{d\ln r} = (1-\Gamma_3)\frac{\mathcal{N}}{\mathcal{D}} + (1-\Gamma_3)\mathcal{B} + \frac{2\pi r^2(F^+ - F^-)}{\dot{M}c_V T}. \tag{88}$$

There are two singular points of these equations: horizon $r = r_H$, defined by $\Delta(r_H) = 0$, and sonic point $r = r_s$ defined by the condition

$$V^2(r_s) = \frac{\mathcal{C}(r_s)c_s^2(r_s)}{1 + 2c_s^2\left(1-\beta^*\right)/(1+\beta^*)}. \tag{89}$$

(i) *The horizon.* At the horizon $\Delta = 0$ and therefore $\mathcal{B} = \infty$, $\mathcal{A} = \infty$. However, one may demonstrate that $\mathcal{N} = 1$ and $\mathcal{A}/\mathcal{B} = -1$. Since on the horizon $V^2(r_H) = 1$, it follows that at the horizon $dV/dr$ is non-singular. For $\beta^* \neq 1$ the gradients $d\Sigma/dr$, $dP/dr$ and $dT/dr$ are singular only through the infinit boost required by the hydrostatic equilibrium equation (44). Note, that from equation (34) it follows that at the horizon $\tilde\Omega = 0$. There are no extra regularity conditions at the horizon.

(ii) *The sonic point.* For large radii the flow is subsonic, $V^2 < c_s^2$, and as we have just seen, it crosses the horizon with the light speed, $1 = V^2 > c_s^2$. Thus, somewhere in the flow it must be $V^2 = c_s^2$, i.e. the flow must pass through the sonic point. Derivatives $dV/dr$ and $dT/dr$ are non-singular at the sonic point only if,

$$N = D = 0. \tag{90}$$

This gives an extra algebraic condition which must be satisfied by an acceptable, regular at the sonic point, global solution of the slim disk equations. The regularity condition at the sonic point was the main source of numerical difficulties in solving the PW slim disk equations: only for one particular choice of the eigenvalue $\mathcal{L}_0$ the condition (90) could be satisfied.

One may demonstrate that in the limit of spherical accretion ($\Omega = 0$, $\mathcal{N} = 1$) onto a non-rotating black hole ($a = 0$), the condition (90) becomes (for $\beta^* = 1$),

$$V^2(r_S) = \frac{M}{2\left(r_S - \frac{3}{2}M\right)}, \tag{91}$$

which in the notation of Shapiro & Teukolski (1983) takes the form

$$u^2(r_S) = \frac{M}{2r_S}, \tag{92}$$

that is identical with their formula (G.17).

On the other hand using Eq. (44) in the form

$$H^2 = c_s^2 \mathcal{G} \tag{93}$$

and including into the Eq. (81) the $dH/dr$ term:

$$\mathcal{B} = -\frac{d}{dr}\left(\log \Delta^{1/2} H\right) = -\frac{r - M}{\Delta} - \frac{1}{r}\mathcal{N}, \tag{94}$$

where $\mathcal{N} = d\log H/d\log r$ one obtains from Eqs. (84, 85) (for $\beta^* = 1$) the following condition for the velocity at the sonic point:

$$V(r_S) = \frac{2c_s}{1 + \Gamma_1}, \tag{95}$$

which is the same condition as the one obtained by Narayan, Kato & Honma (1996).

(iii) *The boundary conditions.* The set of equations describing the flow is of the fifth order with four independent radial first derivatives (say, $d\Omega/dr$, $dV/dr$, $d\Sigma/dr$, and $dT/dr$) and one second derivative ($d^2\Omega/dr^2$). Five integration constants are required. The equation of mass conservation is integrated analytically, and this fixes one integration constant, $\dot{M}$, which is a free parameter. Furthermore, since the angular momentum equation has been analytically integrated from the horizon, an (unknown) value of the specific angular momentum at the horizon, $\mathcal{L}_0$, appears in the angular momentum equation. This constant however is not a free parameter and is determined such that the flow is transonic. The two regularity conditions ($N = 0, D = 0$) at the sonic point provide one condition for the sonic point location, and one extra condition for the five integration constants. Three boundary conditions are assumed at the outer boundary. There are no other boundary assumed conditions, in particular no other boundary conditions at the horizon. The solution is determined by $\dot{M}$, $\alpha$, and $M$, wheares $\mathcal{L}_0$, is an eigenvalue that should be determined. In the numerical procedure, we use the self-similar solution (of $\Omega$, $V$, and $T$) as the outer boundary conditions, and by varying $\mathcal{L}_0$, we find its particular value (the eigenvalue of the problem) that fulfils the regularity conditions ($N = 0, D = 0$) (for detail see Chen & Taam 1993).

## 7. NUMERICAL SOLUTIONS

The set of Kerr black hole disk equations was solved by using the method described in Chen & Taam (1993). This method has been modified to take into account the terms introduced by general relativistic effects. Here we focus only on the case of advection dominated accretion flows (ADAFs) and study the effects of the rotation of the black hole. Specifically, we assume the gas is optically thin ($\tau = 0$) and consider the $\beta^* = 1$ case.. The local radiative cooling only through non-relativistic bremsstrahlung emission (Eq. (65). The corresponding solution in pseudo-Newtonian potential approximation was obtained by Chen, Abramowicz, & Lasota (1996) and Narayan, Kato, & Honma (1996).

The mass accretion rate is measured in the Eddington limit which is defined as

$$\dot{M}_E = 4\pi GM/(c\kappa_{es}), \quad \kappa_{es} = 0.34. \tag{96}$$

and the units for $r$ and angular momentum (such as $\mathcal{L}$, $a$) are $r_G$ and $M$ respectively.

We present in Figure 1 examples of solutions with three different rotation rates of the black hole, $a/M = 0$ (solid-line), 0.5 (dotted-line), and 0.99 (dashed-line). The black hole mass is $M = 10{\rm M}_\odot$, the viscosity parameter is $\alpha = 0.1$ and the accretion rate is $\dot{M}/\dot{M}_E = 10^{-5}$. The pseudo-Newtonian solution is also plotted for comparison (heavy dots). One can see that the pseudo-Newtonian solution is an excellent approximation of the solution for a Schwarzschild black hole ($a = 0$). As the black hole rotates faster, the sonic point moves inwards, mainly due to the increase of the sound speed. This is due to the deepening of the potential well depth with increasing black hole rotation since the ADAF temperature is close to the virial one. The specific angular momentum lost to the black hole is smaller for faster rotating black hole which is also the effect of the hole becoming smaller with increasing rotation. At the black hole surface the pressure goes to infinity due to the infinite boost required by Eq. (44).

## 8. CONCLUSION

We have derived a complete set of equations describing accretion flows with non-zero angular momentum around a rotating black hole (non-Keplerian accretion disks). We used the Novikov and Thorne form of the Kerr metric, which is accurate up to the $(H/r)^0$ terms. Since ADAFs can be heated up to the virial temperature and therefore $H/r \lesssim 1$, the slim disk approximation may appear to be not adequate. However, it is known from the study of ADAFs in the Newtonian case (Narayan & Yi 1995a) that the quasi-spherical structure of ADAFs can be well represented by suitably interpreted vertically averaged quantities. We don't expect general relativistic effects to modify this conclusion.

We obtained several test solutions which show that ADAFs in a Kerr space time have the same basic properties as the pseudo-Newtonian ADAFs. The rotation of the hole allows however for higher temperatures and pressures which should modify the observed emission from ADAFs. This property as well as the influence of the general relativistic effects on light propagation will be studied in a future work.

MAA acknowledges a partial support from Nordita through the Nordic Project on Non-Linear phenomena in accretion disks around black holes.

Fig. 1.— The radial structure of the Mach number ($\mathcal{M}$), the pressure ($p$ in cgs unit), the angular momentum ($\mathcal{L}$ in unit of $M$), and the sound speed ($c_s$ in unit of the speed of light). Here the mass of the black hole is $10 M_\odot$, $\alpha = 0.1$ and $\dot{M}/\dot{M}_E = 10^{-5}$. The solid, dotted, and dashed lines represent the cases of $a/M = 0$, 0.5, and 0.99 respectively. The heavy dots represent solutions obtained with the pseudo-Newtonian potential. These solutions are excellent approximation to the solutions representing the Schwarzschild black hole flows (in the case $a = 0$). The corresponding Keplerian angular momenta of test particles around Kerr black holes are also shown for comparison (the thin lines).